\documentclass{ws-ijmpe}

\usepackage{bm}

\newcommand{\be}{\begin{equation}}
\newcommand{\ee}{\end{equation}}
\newcommand{\ben}{\begin{eqnarray}}
\newcommand{\een}{\end{eqnarray}}
\newcommand{\ba}{\begin{array}}
\newcommand{\ea}{\end{array}}
\newcommand{\bc}{\begin{center}}
\newcommand{\ec}{\end{center}}
\newcommand{\bml}{\begin{mathletters}}
\newcommand{\eml}{\end{mathletters}}

\begin{document}

\markboth{W. Satu{\l}a, J. Dobaczewski, P. Olbratowski, M. Rafalski,
T.R. Werner, R.A. Wyss, and M. Zalewski}
{Global nuclear structure aspects of tensor
  interaction}

%
\catchline{}{}{}{}{}
%

\author{W. SATU{\L}A$^{a),b)}$\footnote{satula@fuw.edu.pl},~~M.
ZALEWSKI$^{a)}$,~~J. DOBACZEWSKI$^{a),c)}$,~~P. OLBRATOWSKI$^{a)}$, \\
M. RAFALSKI$^{a)}$,~~T.R. WERNER$^{a)}$, and~~R.A. WYSS$^{b)}$ }

\address{$^{a)}$Institute of Theoretical Physics, University of Warsaw,
ul. Ho\.za 69, 00-681 Warsaw, Poland \\
$^{b)}$ KTH (Royal Institute of Technology),
AlbaNova University Center, 106 91 Stockholm, Sweden \\
$^{c)}$Department of Physics, P.O. Box 35 (YFL),
FI-40014 University of Jyv\"askyl\"a, Finland
}

\title{GLOBAL NUCLEAR STRUCTURE ASPECTS OF TENSOR
  INTERACTION}

\maketitle
\begin{history}
\received{(received date )}
\revised{(revised date )}
\end{history}

\begin{abstract}
A direct fit of the isoscalar spin-orbit
and both isoscalar and isovector tensor coupling constants to
the $f_{5/2}-f_{7/2}$ SO splittings in $^{40}$Ca, $^{56}$Ni, and $^{48}$Ca
requires ({\it i\/}) a significant reduction of the standard isoscalar spin-orbit strength
and ({\it ii\/}) strong attractive tensor coupling constants.
The aim of this paper is to address the consequences of these strong
attractive tensor and weak spin-orbit fields on total binding energies,
two-neutron separation energies and nuclear
deformability.
\end{abstract}

\section{Introduction}

Density functional theory (DFT) is a method of choice in
large-scale calculations of nuclear properties.
In spite of certain difficulties related to rigorous formulation of the DFT for
self-bound systems like atomic nuclei, the method is
potentially exact, which is guaranteed by the Hohenberg-Kohn-Scham (HKS)
theorems\cite{[Hoh64],[Koh65a]}, see recent discussion in
Refs.\cite{[Eng07],[Gir08],[Gir08a]}.

Due to the complexity of nuclear many-body problem in general, and
of the effective nucleon-nucleon interaction in particular,
there exist no universal rules for constructing
nuclear energy density functional (EDF). In this
respect, our intuition is almost solely based on symmetry properties and
practical knowledge, accumulated over the years for
density-dependent effective interactions of Skyrme\cite{[Sky56xw]} or
Gogny\cite{[Gog75]} type being applied within the mean-field  (MF)
approximation. Free parameters of these interactions
or functionals are fitted to empirical data.
Hence, the quality and performance of these methods strongly depends
on adopted fitting strategies and datasets\cite{[Klu08]}.

Conventional fitting methods use
datasets that are dominated by bulk nuclear matter data and
by nuclear binding energies of selected double-magic nuclei, with
essentially no data on the single-particle (s.p.) energies.
Such strategies have quite dramatic consequences concerning mostly
spin-orbit (SO) and tensor parts of the EDF.
In particular, they lead to an artificial isoscalar-effective-mass
scaling of the SO strengths\cite{[Zdu05xws]},
contradicting scaling in selected s.p.\ splittings\cite{[Sat08]}, and
perpetual problems in reproducing evolution of the proton (neutron) s.p.\
energy splittings versus the neutron (proton) shell filling along the
isotopic or isotonic chains of nuclei.

The most prominent
examples of such chains include: neutron-rich oxygen\cite{[Bec06s]},
neon\cite{[Bel05s]}, sodium\cite{[Uts04s],[Tri05s]},
magnesium\cite{[Ney05s]}, titanium\cite{[For04w],[Din05w]}, or, in
medium-mass region, antimony\cite{[Sch04s]} isotopes. In fact, a
non-conventional shell evolution found in these neutron-rich nuclei
directly motivated the shell-model theorists to introduce the so-called
monopole shifts, to account for empirical trends.
The physical origin of these shifts was, in turn, attributed to the
shell-model tensor interaction\cite{[Ots01s],[Ots05s],[Hon05s]}.
Connection between the monopole-shifts and the tensor interaction
was later on confirmed within the self-consistent MF models
using either finite-range Gogny force\cite{[Ots06]} or contact Skyrme
interaction\cite{[Dob06s],[Bro06],[Col07],[Bri07w],[Gra07],[Zal08],[Zhu08]}
augmented by a strong tensor interaction.

Single-particle spectra provide for a clear evidence of
strong tensor interaction and call for a
new strategy of fitting the nuclear EDF in general, and of the SO
and tensor terms in particular, directly to the s.p.\
data\cite{[Zal08],[Kor08]}.
The use of the s.p.\ levels was usually contested,
because of the isoscalar-effective-mass scaling ($m^*$)
of s.p.\ levels. Several authors\cite{[Ham76],[Ber80w],[Lit06w]} argued
that the physical density of s.p.\ levels around the Fermi
energy can be reinstated only after the inclusion of particle-vibration
coupling, that is, by going beyond MF. In our opinion,
effective EDF theories should warrant a proper value of the effective
mass through the fit to empirical data
and readjust other coupling constants
to this particular value of $m^*$, leading to fairly  $m^*$ independent
predictions. Hence, fitting strategies can include information on s.p.\ levels
provided that the s.p.\ levels are understood through the binding-energy differences
between doubly-magic cores and the lowest s.p.\ states in
odd-$A$ single-particle/hole neighbors\cite{[Rut99a],[Zal08]}. Spherical s.p.\ energies or, more
precisely, the Kohn-Sham s.p.\ energies computed in even-even double-magic core
should serve only as auxiliary quantities.

In our recent study\cite{[Zal08]}, we have proposed a
novel fitting strategy of the SO and tensor terms in
the nuclear EDF. It is based on a direct fit to
the $f_{7/2} - f_{5/2}$ SO splittings in spin-saturated isoscalar nucleus
$^{40}$Ca, spin-unsaturated isoscalar nucleus $^{56}$Ni, and spin-unsaturated
isovector nucleus $^{48}$Ca. The procedure allows for fixing three out of four
coupling constants in this sector, namely, the isoscalar strengths
of the SO and tensor interactions and the ratio of the isovector
coupling constants. The procedure indicates a clear need for a major
reduction of the SO strength and for strong attractive tensor fields.
The aim of the present work is to address further consequences of strong
attractive tensor and weak SO fields on binding energies, two-neutron
separation energies, and nuclear
deformability.

\section{Fitting the tensor strengths to single-particle energies}

\begin{figure}[t]
\centering
\includegraphics[width=11.5cm,clip]{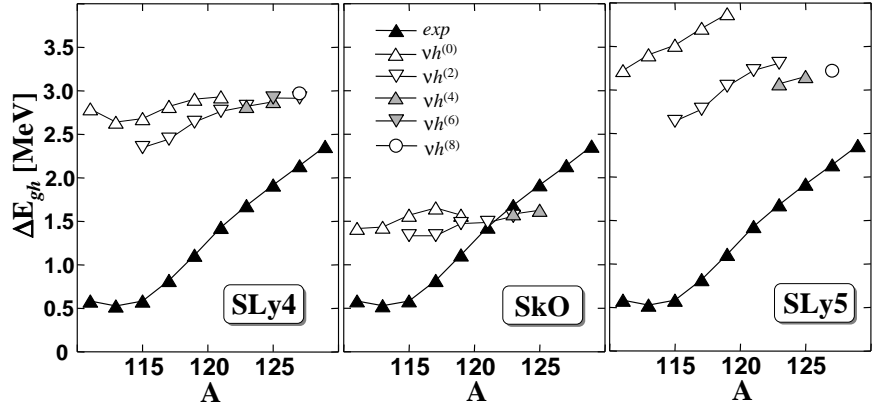}
\caption[a]{The $\pi h_{11/2}-\pi g_{7/2}$ splitting in
$_{51}^A$Sn isotopes versus $A$. Black triangles label
empirical data taken from Ref.~\protect\cite{[Sch04s]}. Open and gray
symbols represent the SHF results obtained by using the SLy4 (left)
SkO (middle), and SLy5 (right) parameterizations, respectively.
Different symbols labeling theoretical results
follow the SHF minima corresponding to configurations differing in
the $\nu h_{11/2}$ occupancies as indicated in the legend.
\label{fig1}}
\end{figure}

The s.p.\ levels constitute one of the main building blocks of the MF method. In spite
of that, the Skyrme HF (SHF) method that uses forces fitted to
bulk nuclear properties performs rather poorly with regard to
the s.p.\ SO splittings\cite{[Zal08],[Kor08]}. This is visualized in
Fig.~\ref{fig1}, showing the $\pi 1h_{11/2}-\pi 1g_{7/2}$ splittings,
$\Delta E_{gh}$, in antimony, calculated by using the
SLy4\cite{[Cha97fw]}, SkO\cite{[Rei99]}, and SLy5\cite{[Cha97fw]}
parameterizations. The SLy4 force strongly overestimates the
absolute value of  $\Delta E_{gh}$ and fails to reproduce the
slope of the $\Delta E_{gh} (A)$ curve. The non-standard isovector SO in the
SkO force helps by reducing, on average, the splitting to the empirical level, but
does not change the slope of the $\Delta E_{gh} (A)$ curve. Finally, in SLy5,
the inclusion of
tensor terms changes the slope, but shifts the theoretical curves in a wrong
direction. The latter observation suggests that the fit to masses leads to
values of tensor
coupling constants that are at variance with those deduced from the s.p.\
level analysis, see Refs.\cite{[Les07],[Zal08]}. However, one should point out
that the $\pi 1h_{11/2}-\pi
1g_{7/2}$ splittings depend upon many factors including, apart from the SO and
tensor fields, the effective mass, centroid energies of the $\ell = 4$ and
$\ell = 5$ sub-shells, and strong polarization effects.  Hence, conclusions
concerning the SO and tensor coupling constants that are deduced solely from
these data should be considered to be tentative.

\begin{figure}[t]
\centering
\includegraphics[width=12cm,clip]{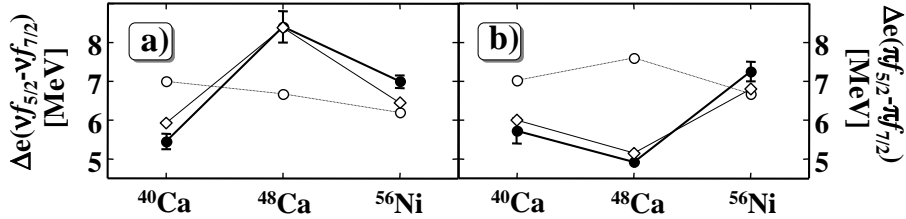}
\caption[a]{The neutron (top) and proton
(bottom) $1f_{7/2}-1f_{5/2}$ SO splittings in
$^{40}$Ca, $^{48}$Ca, and $^{56}$Ni. Black symbols show the
mean empirical values taken from Refs.\cite{[Oro96w],[Sch07a]} and
open dots denote the SkO results. Open diamonds represent the
results obtained by using the SkO$_{T^\prime}$ functional
of Ref.\cite{[Zal08as]}, which includes strong attractive tensor
terms and a reduced SO strength.
\label{fig2}}
\end{figure}

It is well known, see Refs.\cite{[Flo75],[Bri77]}, that
the tensor interaction strongly modifies the SO one-body potential.
In the spherical-symmetry limit, the isoscalar ($t=0$) and isovector ($t=1$)
SO one-body potentials read:
\begin{eqnarray}\label{sot} W_t^{SO} &
= & \frac{1}{2r}\left( C^J_t J_t(r) - C^{\nabla J}_t \frac{d\rho_t}{dr}\right)
{\mathbf L} \cdot {\mathbf S},
\end{eqnarray}
where $C^J_t$ and  $C^{\nabla J}_t$ are the tensorial and spin-orbit
coupling constants, see for example Ref.\cite{[Zal08]}. The tensor field
depends upon the radial component of the spin-orbit vector density ${\mathbf
J}_t=\frac{{\mathbf r}}{r}J_t(r)$ that measures the
spin-asymmetry of the nucleus and can rapidly vary with particle numbers. On the contrary,
the second term in Eq.~(\ref{sot}), which is due to the conventional two-body
spin-orbit interaction, depends on the radial
derivative of the particle density $\rho_t$, which varies relatively slowly
with particle numbers. Such a contrasting behavior of the two major constituents of
the SO potential can be actually used  to fit the
coupling constants to data\cite{[Zal08]}. The idea is visualized in
Fig.~\ref{fig2}, which shows the $1f_{7/2}-1f_{5/2}$ SO splittings in $^{40}$Ca,
$^{48}$Ca, and $^{56}$Ni.  These splittings form a very distinct pattern
that cannot be reproduced based solely on the conventional SO potential.
Indeed, the $1f_{7/2}-1f_{5/2}$ SO splittings in $^{40}$Ca,
$^{48}$Ca, and $^{56}$Ni are fairly constant when calculated using,
for example the SkO force, see curve marked by open dots in
Fig.~\ref{fig2}. It reflects the fact that
the neutron and proton radial form-factors  $\frac{d\rho}{dr}$ almost do not
change when going from $^{40}$Ca through $^{48}$Ca to $^{56}$Ni.
At the same time the neutron and proton SO vector
densities $J(r)$ change rapidly when
going from the isoscalar spin-saturated  $^{40}$Ca to the isoscalar
spin-unsaturated nucleus $^{56}$Ni, and, finally, to the isovector spin-unsaturated nucleus $^{48}$Ca.
This allows for a simple and intuitive three-step fitting procedure\cite{[Zal08]} of
the  $C^{\nabla J}_0$ in $^{40}$Ca, $C^{J}_0$ in $^{56}$Ni, and
$C^{J}_1/ C^{\nabla J}_1$ ratio in $^{48}$Ca.
This procedure leads to ({\it i\/}) a significant reduction in the isoscalar
SO strength and ({\it ii\/}) strong attractive tensor coupling constants.
It systematically
improves such s.p.\ properties as the SO splittings
and magic-gap energies\cite{[Zal08]}, but leads to deteriorated nuclear binding energies.

\section{Tensor interaction and the nuclear binding
energies}\label{sec3}

\begin{figure}[t]
\centering
\includegraphics[width=8cm,clip]{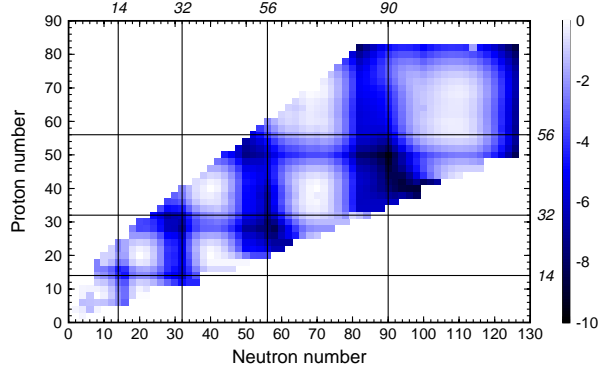}
\caption[a]{Tensor contributions to the total binding energy calculated
by using the spherical Hartree-Fock-Bogoliubov model with the SLy4$_T$
functional of Ref.\protect\cite{[Zal08]}. Vertical and horizontal lines indicate the tensorial
magic numbers. From Ref.\protect\cite{[Zal08b]}.
\label{fig3}}
\end{figure}

\begin{figure}[t]
\centering
\includegraphics[width=6cm,clip]{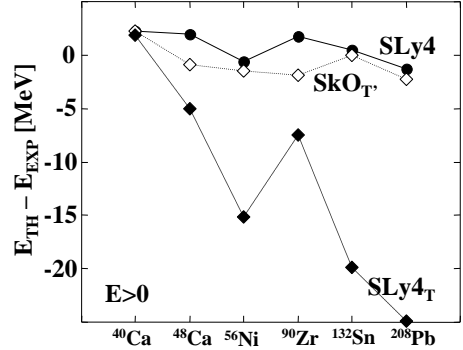}
\caption[a]{Differences between theoretical and experimental binding energies
(positive) in selected double-magic nuclei. Black dots represent the results
obtained using conventional SLy4 force. Black and white diamonds label the
results calculated using the SLy4$_T$ and the
SkO$_{T^\prime}$ functionals, respectively.
\label{fig4}}
\end{figure}

As discussed in Ref.\cite{[Zal08b]}, the tensor contribution to the nuclear binding
energy shows interesting generic topological patterns
closely resembling those of the shell-correction, see Fig.~\ref{fig3}.
The single-particle
{\it tensorial magic numbers\/} at $N(Z)$=14, 32, 56, or
90, corresponding to the maximum spin-asymmetry in the $1d_{5/2}$,
$1f_{7/2}\oplus 2p_{3/2}$,  $1g_{9/2}\oplus 2d_{5/2}$ and $1h_{11/2}\oplus
2f_{7/2}$ spherical s.p.\ configurations, respectively,
are clearly seen in the figure.  Note, that the calculated
tensorial magic numbers are shifted due to configuration mixing toward the classic
magic numbers of $N(Z)$=8, 20, 28, 50, and 82.
The topological features shown in
Fig.~\ref{fig3} are fairly independent of a specific parameterization of the force.
Indeed, they simply reflect the order of s.p.\ levels, which is
rather unambiguously established  and relatively well reproduced by the
state-of-the art nuclear MF models, at least in light and medium-mass
nuclei.

Values of the tensor and SO strengths deduced from the s.p.\ properties are at variance
with those obtained from mass fits\cite{[Les07],[Zal08as]}.
A large reduction of the SO strength, which is particularly strong for
the low-$m^*$ forces like SLy4, has a particularly
destructive impact on theoretical binding energies, see Fig.~\ref{fig4}.

A multidimensional fit to masses shows that the mass performance of
the SLy4$_T$ force can be improved by a tiny refinements of the remaining
coupling constants\cite{[Zal08]}, however, the accuracy of the
original SLy4 cannot be regained. This indicates that a {\it spectroscopic quality\/}
parameterization that would perform reasonably well on binding energies must have
large effective mass, $m^*$ $\geq 0.9$.  One of the candidates is the SkO$_{T^\prime}$
functional of  Ref.\cite{[Zal08as]}. This functional, at least for
the classical set of double-magic nuclei shown in Fig.~\ref{fig4},
is of a similar accuracy as SLy4, and it outperforms both
the SLy4$_T$ and SLy4$_{T\mathrm{min}}$ of Ref.\cite{[Zal08]}.

\begin{figure}[t]
\centering
\includegraphics[width=6cm,clip]{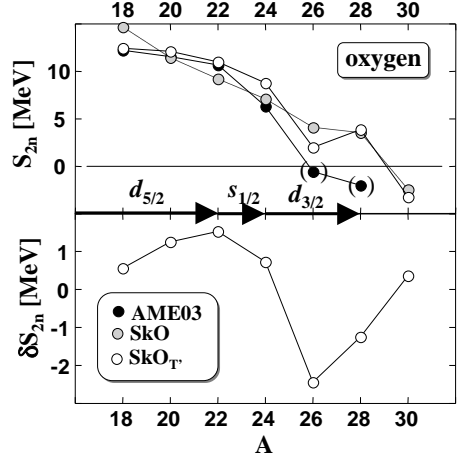}
\caption[a]{Two-neutron separation energy $S_{2n}$ (upper part) in
oxygen isotopes. Empirical data~\cite{[Aud03]} are labeled by black
dots. Theoretical values obtained using the SkO and SkO$_{T^\prime}$
functionals are marked by gray and white dots, respectively.
Lower part shows contribution of the tensor term to the $S_{2n}$ as
a function of shell filling.
\label{fig5}}
\end{figure}

This allows for reasonable quantitative estimates of the tensor influence,
for example, on two-neutron
separation energies, potential energy surfaces (PES's), and
onset of deformation.  An example of
calculation of the two-neutron
separation energies for oxygen nuclei is shown in Fig.~\ref{fig5}. One clearly sees
here the way the tensor interaction induces in oxygen isotopes a breaking of stability against
the two-neutron emission around $^{26}$O. Indeed, as shown in the lower panel
of the figure, in $^{26}$O a decrease of $S_{2n}$ is directly related to
the $d_{3/2}$ sub-shell occupation that reduces the spin-asymmetry and tensor
contribution to the binding energy.

\begin{figure}[t]
\centering
\includegraphics[width=6cm,clip]{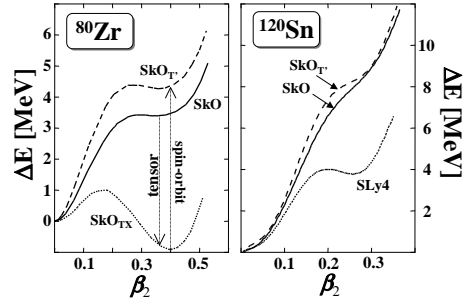}
\caption[a]{Potential energies versus axial quadrupole deformation
parameter $\beta_2$ in $^{80}$Zr (left) and $^{120}$Sn (right),
calculated by using the SkO,
SkO$_{TX}$, SkO$_{T^\prime}$, and  SLy4 functionals. See text for
details.
\label{fig6}}
\end{figure}

By deforming the nucleus one can easily change the spin asymmetry and, in turn,
tensor effects. Fig.~\ref{fig6} shows the PES's
versus quadrupole deformation in $^{80}$Zr (left) and  $^{120}$Sn (right),
calculated by using the quadrupole-constrained HFB method.
At the spherical shape, nucleus $^{80}$Zr is spin-saturated. By deforming the system,
one increases the spin-asymmetry by enforcing the occupation of the $g_{9/2}$
sub-shell. By adding to SkO a strong attractive tensor field (SkO$_{TX}$), one
pulls the deformed minimum down. The consecutive reduction of the SO
strength (SkO$_{T^\prime}$) provides for a compensation mechanism, and it shifts the
$g_{9/2}$ sub-shell and the deformed minimum up in energy, back to its original
position obtained for the SkO functional. In $^{56}$Ni, a similar compensation mechanism
is found in the yrast super-deformed bands, see
Ref.\cite{[Zal08as]}. In case of $^{120}$Sn, the PES's
calculated by using the SkO and SkO$_{T^\prime}$ parameterizations are again very close to each
other. Note however, that both these curves differ substantially from the PES
calculated using the SLy4 parameterization.

\section{Summary}

In this study, we discussed specific nuclear-structure effects induced by strong attractive tensor fields
and weak spin-orbit field, which result from direct fits of the coupling constants to the SO splittings.
In particular, we showed that
contributions to the nuclear binding energies that are due to the tensor field show
a generic {\it magic structure\/} with {\it tensorial magic numbers\/} at
$N(Z)$=14, 32, 56, or 90 corresponding to maximum spin-asymmetry in
$1d_{5/2}$, $1f_{7/2}\oplus 2p_{3/2}$,  $1g_{9/2}\oplus 2d_{5/2}$ and
$1h_{11/2}\oplus 2f_{7/2}$ single-particle configurations.
We also demonstrated that it is possible to construct
a functional being able to simultaneously reproduce the $f_{5/2}-f_{7/2}$
SO splittings in $^{40}$Ca, $^{56}$Ni, and $^{48}$Ca nuclei and binding
energies of doubly magic spherical nuclei. By using this particular functional,
we discussed specific structural effects,  pertaining to strong tensor
terms, exerted on
two-neutron separation energies in oxygen isotopes and
PES's in $^{80}$Zr and $^{120}$Sn.
In particular, in the context of nuclear deformation properties,
we discussed compensation mechanism
between the attractive tensor fields and weak SO field.


This work was supported in part by the Polish Ministry of
Science under Contract No.~N~N202~328234,
by the Academy of Finland and University of
Jyv\"askyl\"a within the FIDIPRO programme, and by the
Swedish Research Council.


\end{document}